\catcode`\@=11
\newif\if@fewtab\@fewtabtrue

{\count255=\time\divide\count255 by 60
\xdef\hourmin{\number\count255}
\multiply\count255 by-60\advance\count255 by\time
\xdef\hourmin{\hourmin:\ifnum\count255<10 0\fi\the\count255}}
\def\ps@draft{\let\@mkboth\@gobbletwo
    \def\@oddhead{}
    \def\@oddfoot
       {\hbox to 7 cm{$\scriptstyle Draft\ version:\ \draftdate$
       \hfil}\hskip -7cm\hfil\rm\thepage \hfil}
    \def\@evenhead{}\let\@evenfoot\@oddfoot}

\def\ceqno{\global\@fewtabfalse
    \ifcase\@eqcnt \def\@tempa{& & &}\or \def\@tempa{& &}
      \or \def\@tempa{&}
      \or\def\@tempa{}\fi\@tempa
{\rm(\theequation)}}

\def\aeqno#1{\global\@fewtabfalse
    \ifcase\@eqcnt \def\@tempa{& & &}\or \def\@tempa{& &}
      \or \def\@tempa{&}
      \or\def\@tempa{}\fi\@tempa
{\rm(\theequation,#1)}}

\def\label#1{\ifnum\draftcontrol=1
 \global\def\draftnote{$\scriptstyle #1$}\fi
 \@bsphack\if@filesw {\let\thepage\relax
   \def\protect{\noexpand\noexpand\noexpand}%
\xdef\@gtempa{\write\@auxout{\string
      \newlabel{#1}{{\@currentlabel}{\thepage}}}}}\@gtempa
   \if@nobreak \ifvmode\nobreak\fi\fi\fi
  \@esphack}

\def\alabel#1#2{\label{#1}\global\@fewtabfalse
    \ifcase\@eqcnt \def\@tempa{& & &}\or \def\@tempa{& &}
      \or \def\@tempa{&}
      \or\def\@tempa{}\fi\@tempa
{\hbox to 3cm{\phantom{\rm(\theequation,#2)}
\draftnote \hfil}\hskip -3cm {\rm(\theequation,#2)}}}

\def\clabel#1{\label{#1}\global\@fewtabfalse
    \ifcase\@eqcnt \def\@tempa{& & &}\or \def\@tempa{& &}
      \or \def\@tempa{&}
      \or\def\@tempa{}\fi\@tempa
{\hbox to 3cm{\phantom{\rm(\theequation)}
\draftnote \hfil}\hskip -3cm{\rm(\theequation)}}}

\def\eqnarray{\def\draftnote{{}}\global\@fewtabtrue
\stepcounter{equation}\let\@currentlabel=\theequation
\global\@eqnswtrue
\global\@eqcnt\z@\tabskip\@centering\let\\=\@eqncr
$$\halign to \displaywidth\bgroup\@eqnsel\hskip\@centering\@eqcnt\z@
  $\displaystyle\tabskip\z@{##}$&\global\@eqcnt\@ne
  \hskip 1\arraycolsep \hfil${##}$\hfil
  &\global\@eqcnt\tw@ \hskip 1\arraycolsep
$\displaystyle\tabskip\z@{##}$
\hfil  \tabskip\@centering&\global\@eqcnt\thr@@\llap{##}\tabskip\z@
\cr}

\def\endeqnarray{\@@eqncr\egroup
      \global\advance\c@equation\m@ne$$\global\@ignoretrue}

\def\@eqnnum{\hbox to 3cm{\phantom{\rm(\theequation)} \draftnote
                         \hfil}\hskip -3cm {\rm(\theequation)}}

\def\@@eqncr{\let\@tempa\relax
    \ifcase\@eqcnt \def\@tempa{& & &}\or \def\@tempa{& &}
      \or \def\@tempa{&}
      \or\def\@tempa{}
\fi\@tempa
\if@eqnsw
\if@fewtab\@eqnnum\fi
\stepcounter{equation}\fi\global
\@eqnswtrue\global\@eqcnt\z@\global\@fewtabtrue\cr}

\def\draftcite#1{\ifnum\draftcontrol=1#1\else{}\fi}

\def\@lbibitem[#1]#2{\item{}\hskip -3cm \hbox to 2cm
{\hfil$\scriptstyle\draftcite{#2}$}\hskip
1cm[\@biblabel{#1}]\if@filesw
     {\def\protect##1{\string ##1\space}\immediate
      \write\@auxout{\string\bibcite{#2}{#1}}}\fi\ignorespaces}

\def\@bibitem#1{\item\hskip -3cm \hbox to 2cm
{\hfil $\scriptstyle\draftcite{#1}$}\hskip 1cm
\if@filesw \immediate\write\@auxout
       {\string\bibcite{#1}{\the\value{\@listctr}}}\fi\ignorespaces}

\def\nsection#1{\section{#1}}

\font\tendl=msbm10  scaled \magstep1
\font\sevendl=msbm7 scaled \magstep1
\font\fivedl=msbm5 scaled \magstep1
\font\tengl=eufm10  scaled \magstep1
\font\sevengl=eufm7 scaled \magstep1
\font\fivegl=eufm5 scaled \magstep1

\newfam\dlfam \def\dl{\fam\dlfam\tendl} 
\textfont\dlfam=\tendl \scriptfont\dlfam=\sevendl
\scriptscriptfont\dlfam=\fivedl
\newfam\glfam \def\gl{\fam\glfam\tengl} 
\textfont\glfam=\tengl \scriptfont\glfam=\sevengl
\scriptscriptfont\glfam=\fivegl

\def\draftdate{\number\month/\number\day/\number\year\ \ \ \hourmin }

\global\def\draftcontrol{0}
\catcode`\@=12
\def\tilde{\widetilde}
\def\hat{\widehat}

\def\theequation{\arabic{equation}} 

\newcommand{\be}{\begin{eqnarray}}
\newcommand{\en}{\end{eqnarray}\vs 0.5 cm}

\newcommand{\vs}{\vskip}

\newcommand{\NR}{{{\dl R}}}
\newcommand{\NC}{{{\dl C}}}
\newcommand{\NZ}{{{\dl Z}}}
\newcommand{\qq}{\begin{eqnarray}}

\newcommand{\ee}{{\rm e}}

\newcommand{\qqq}{\end{eqnarray}}

\newcommand{\tr}{\hbox{tr}}
\newcommand{\ad}{\hbox{ad}}

\newcommand{\CA}{{\cal A}}

\newcommand{\CC}{{\cal C}}

\newcommand{\CF}{{\cal F}}
\newcommand{\CG}{{\cal G}}
\newcommand{\CH}{{\cal H}}

\newcommand{\CL}{{\cal L}}

\newcommand{\CP}{{\cal P}}

\newcommand{\CU}{{\cal U}}

\newcommand{\CX}{{\cal X}}

\documentstyle[11pt]{article}
\newcommand{\parr}{\partial_{r}\, } 
\newcommand{\parl}{\partial_{l}\, } 
\newcommand{\AL}{A_{_l}} 
\newcommand{\AR}{A_{_r}} 
\newcommand{\Ad}{{\rm Ad}} 
\newcommand{\gp}{{g_{_+}}} 
\newcommand{\gm}{{g_{_-}}} 
\newcommand{\gpinv}{{g_{_{+}}^{-1}}}
\newcommand{\gminv}{{g_{_{-}}^{-1}}} 
\newcommand{\ggl}{{\gl g}} 
\newcommand{\jlp}{{\gpinv\parl\gp}} 
\newcommand{\jrp}{{\gpinv\parr\gp}}
\newcommand{\jlm}{{\gminv\parl\gm}} 
\newcommand{\jrm}{{\gminv\parr\gm}}
\begin{document}

\title{Boundary $\,G/G\,$ theory\\
and\\ topological Poisson-Lie sigma model}

\author{\ \\Fernando Falceto \\ Depto. F\'{\i}sica
Te\'orica, Univ. Zaragoza, \\ E-50009 Zaragoza, Spain
\\  \\Krzysztof Gaw\c{e}dzki \\ C.N.R.S., I.H.E.S., 
F-91440  Bures-sur-Yvette, France\\ 
and Laboratoire de Physique, 
ENS-Lyon,\\46, All\'ee d'Italie, F-69364 Lyon, France}
\date{ }
\maketitle

\vskip 0.3cm
\begin{center}
\end{center}
\vskip 1 cm

\begin{abstract}
\vskip 0.3cm
\noindent We study a boundary version of the gauged WZW model 
with a Poisson-Lie group $G$ as the target. The Poisson-Lie structure
of $G$ is used to define the Wess-Zumino term of the action on surfaces
with boundary. We clarify the relation of the model to the topological 
Poisson sigma model with the dual Poisson-Lie group $G^*$ as the target
and show that the phase space of the theory on a strip is essentially 
the Heisenberg double of $G$ introduced by Semenov-Tian-Shansky.

\end{abstract}
\vskip 2cm

\nsection{Introduction}

The topological Poisson sigma models introduced in \cite{SS} 
have become an important laboratory for field-theoretical
methods applied to mathematical problems, see 
\cite{Konts}\cite{cafe}\cite{cafe2}. In those applications,
it is important to consider the models on two-dimensional surfaces
with boundary, for example on a disc. In particular,
the perturbative treatment of the quantum theory in such
a geometry leads to a formal deformation quantization of the
algebra of functions on the target, with the leading order
of the $*$-product of functions determined by the Poisson bracket.
In the specific case when the target manifold is 
a Poisson-Lie group, this approach should provide a royal
route to quantum groups. Although this expectation has not yet
materialized, it remains that the case of topological Poisson-Lie
sigma models deserves a special attention.

It has been observed quite early \cite{AlSchStr} that for the target
space which is the Poisson-Lie group $G^*$ dual to a simple complex 
Lie group $G$ equipped with an $r$-matrix Poisson structure, the 
Poisson sigma model in the bulk is essentially equivalent on the 
classical level to the gauged Wess-Zumino-Witten model, the so called 
$G/G$ coset theory \cite{GawKup}. The latter is a topological field 
theory whose quantum partition and correlation functions on closed 
surfaces compute the Verlinde dimensions of the spaces of non-abelian 
theta functions \cite{Sorger}.

The purpose of this note is four-fold. First, we present 
a covariant description of the relation between the classical 
Poisson-Lie sigma model with $G^*$ target and the $G/G$ coset 
theory, obtained in \cite{AlSchStr} in the Hamiltonian approach.
Second, we extend the discussion to space-times with boundary. 
In the latter case, the requirement of equivalence 
to the Poisson-Lie sigma model fixes the boundary contributions 
in the gauged WZW model. More exactly, the precise equivalence of 
two theories requires a replacement of the target $G^*$ 
in the Poisson sigma model by its quotient by a discrete subgroup
and a restriction of the target in the gauged WZW model
to an open subset of $G$. As a result, the action of the 
$G/G$ theory becomes well defined for any value of the coupling
constant $k$. 

Third, we describe the canonical structure of the resulting classical 
field theory on a cylinder and on a strip. In the first case,
it is essentially the space of conjugacy classes of commuting pairs 
of elements in $G$. In the second case, the phase space appears 
to be a version of the Heisenberg double of Poisson-Lie group 
$G$ introduced by Semenov Tian-Shansky in \cite{S-T-S}, a 
Poisson-Lie replacement for the cotangent bundle $T^*G$. 
This provides an example of a general construction of a Poisson
Lie groupoid of a Poisson manifold described in \cite{cafe2}.

Finally, we discuss briefly the quantization of the theory for
the Poisson structures induced by the  standard 
classical $r$ matrices. The restriction of field values in 
the gauged WZW theory imposed by the equivalence to the topological 
Poisson-Lie theory appears to be not quite innocuous, at
least for generic values of $k$. It leads, for example, to an 
infinite-dimensional space of boundary states, despite a topological 
character of the theory. Nevertheless, as is generally the case 
for two-dimensional topological field theories \cite{Laza}, this 
space has a structure of an associative non-commutative algebra. 
It may be viewed as the quantum deformation $\CU_q(\ggl)$ of the 
enveloping algebra of Lie algebra $\ggl$ of $G$. Another problem 
is the absence of unitarity that has a classical source: the action 
of the gauged WZW model equivalent to the Poisson-Lie sigma 
model is not real in the Minkowski signature with respect to the 
conjugation of fields $g$ and $A$ that fix the compact real forms
of $G$ and $\ggl$. For $q$ a root of unity (i.e. for integer $k$), 
one may reduce the algebra of the boundary states of the $G/G$ 
theory considered here to a finite-dimensional one to recover 
a genuine model of quantum topological field theory.
This is a different version of the boundary $G/G$ theory than
the one constructed recently in \cite{GawGH} within an approach 
respecting unitary at all stages. The gain from the present approach 
is its direct relation to the topological Poisson sigma models and 
Poisson-Lie theory.

The paper is organized as follows. In Sect.\,\,2 we recall
various facts from the theory of Poisson-Lie groups needed later.
Sect.\,\,3 defines a version of the gauged WZW model on surfaces 
with boundary by restricting appropriately the target. Sect.\,\,4 
establishes the relations between the restricted $G/G$ theory and 
the Poisson-Lie sigma models with a discrete quotient of $G^*$
as the target. The phase-space structure of the theory in closed and
open geometries is discussed in Sect.\,\,5 and the remarks about 
quantization of the theory are contained in Sect.\,\,6.

{\bf Acknowledgement} \  F.F. is greatful for hospitality to the IHES 
in Bures-sur-Yvette where the work on this paper was started.

\nsection{Basic Poisson-Lie theory}

Let us start by recalling few basic facts from the theory
of Poisson-Lie groups, see \cite{S-T-S}\cite{LuW}\cite{AlMa}\cite{FG}. 
A Poisson-Lie group is a Lie group equipped with a Poisson structure 
which makes the product $m:G\times G\rightarrow G$ a Poisson map 
if $G\times G$ is considered with the product Poisson structure. 
Linearization of the Poisson structure at the unit $1$ of $G$ 
provides a Lie algebra structure on ${\ggl}^*=T^*_1(G)$ by the 
formula 
\begin{eqnarray}
[d\xi(1),d\zeta(1)]^*=d\{\xi,\zeta\}(1).
\end{eqnarray}
The fact that the product in $G$ is Poisson  
is reflected in the compatibility condition 
\begin{eqnarray}\label{comp} 
\left< [X,Y],[v,w]^*\right>&+&\left<\ad_v^*X,\ad_Y^*w\right>
-\left<\ad_w^*X,\ad_Y^*v\right>\cr
&-&\left<\ad_v^*Y,\ad_X^*w\right>+\left<\ad_w^*Y,\ad_X^*v\right>=0.
\end{eqnarray}
On the other hand if $G$ is connected and simply connected, this is the 
only condition for integrating $[\ ,\ ]^*$ to a Poisson structure
on $G$ that makes it Poisson-Lie and the Poisson structure is unique.
Since $\ggl$ and $\ggl^*$ enter symmetrically in (\ref{comp}),
one has also a connected and simply connected Poisson-Lie group $G^*$ 
with Lie algebra $(\ggl^*,[\ ,\ ]^*)$  and a Poisson 
structure whose linearization at $1$ gives the bracket $[\ ,\ ]$. 
$G^*$ is  the dual Poisson-Lie group of $G$. 

Condition (\ref{comp}) allows to define a Lie bracket in 
$\ggl\oplus\ggl^*$ by the formula
\begin{eqnarray}
[X+v,Y+w]=[X,Y]+[v,w]^*-\ad^*_X w+\ad^*_Yv+\ad^*_wX-\ad^*_vY.
\end{eqnarray}
The corresponding simply connected group $D$, 
together with  its local subgroups $G$ and $G^*$
form a so called local double Lie group \cite{LuW}.
$D$ can be equipped canonically with a Poisson structure,
non-degenerate around the unit,
such that the local embeddings of $G$ and $G^*$
in $D$ are Poisson maps (see \cite{S-T-S}, 
\cite{AlMa} for details). 
$D$ with this Poisson structure is called the Heisenberg double.
It can be considered a generalization of $T^*G$ for Poisson-Lie
groups.
   
The above construction may be described in more concrete terms
that provide an explicit realization for the group $G^*$ and
for the Poisson structures of $G$ and $G^*$. We shall take $G$ 
to be a complex, simple, connected, simply connected Lie group
and we shall work in the complex category. The non-degenerate, 
invariant, bilinear form $\tr$ on $\ggl$ permits to establish an 
isomorphism between $\ggl$ and $\ggl^*$. The Poisson structures will 
be represented by a bivector field $\Gamma$ so that the Poisson 
bracket is given by the contraction of $\Gamma$, i.e. 
$\{\xi,\zeta\}(g)=\iota(\Gamma_g)d\xi\wedge d\zeta$.
For the Poisson structure on $G$, the components of $\Gamma$
contracted with the right-invariant forms $\Lambda(X)=\tr(dgg^{-1}X)$, 
with $X \in {\gl g}$, will be denoted by
$\gamma_g(X ,Y)=\iota(\Gamma_g) \Lambda(X)\wedge\Lambda(Y)$.

As it is shown in ref. \cite{LuW}, for a general 
Poisson-Lie structure on $G$, 
\begin{eqnarray}\label{Gamma}
\gamma_g(X,Y)={1\over 2}\tr( XrY-X Ad_{g}r Ad^{-1}_g Y )
\end{eqnarray}
for an antisymmetric linear operator $r:{\gl g}\rightarrow{\gl g}$
such that
\begin{eqnarray}\label{MYB}
r[rX,Y]+r[X,rY]-[rX,rY]=\alpha [X,Y],\qquad {\rm with}\quad \alpha\in\NC.
\end{eqnarray}
The linearization of the Poisson-Lie structure at the unit $1\in G$
defines a new Lie bracket in $\gl g$, namely
\begin{eqnarray}
[X,Y]_r={1\over 2}\left([X,rY]+[rX,Y]\right).
\end{eqnarray}
The bracket $[\ ,\ ]_r$ in $\gl g$ corresponds to $[\ ,\ ]^*$ in $\gl g^*$, 
via the linear isomorphism induced by the bilinear form $\tr$.

In the following we shall take $\alpha=1$ in (\ref{MYB}) which  
(provided $\alpha\not=0$) 
can be achieved by a trivial rescaling of the bilinear 
structure in $\gl g$ or of the Poisson bracket in $G$.
In this case 
\begin{eqnarray}\label{homo}
[r_\pm X, r_\pm Y]=r_\pm [X,Y]_r
\end{eqnarray}
where $r_\pm={1\over 2}(r\pm I)$
and the embedding
\begin{eqnarray}
\hfill \sigma:\ggl&\rightarrow& {\gl d}=\ggl\oplus\ggl\hfill\cr
\hfill X&\mapsto&(r_+ X,r_-X)
\end{eqnarray}
defines an  homomorphism from $(\ggl, [\ ,\ ]_r)$ 
to $({\gl d} , [\ ,\  ]_{\gl d})$
with  $[\ ,\  ]_{\gl d}=([\ ,\ ], [\ ,\ ])$.
This way one may realize the algebra $(\ggl^*, [\ ,\ ]^*)$ as the 
subalgebra $\ggl_r=\sigma(\ggl)\subset {\gl d}$ and the dual
group $G^*$ as the subgroup $G_r$ in $G\times G$ corresponding 
to the Lie sub-algebra $\ggl_r$. We shall denote by $(g_+,g_-)$
the elements of $G_r$. We may identify $\ggl$ with $\ggl_r^*$ 
using $\sigma$ and $\tr$. Explicitly, the pairing between 
$(r_+X,r_-X)\in\ggl_r$ and $Y\in\ggl$ is given by
\begin{eqnarray}\label{pair}
\tr XY=\tr (r_+X-r_-X)Y.
\end{eqnarray} 
The original bracket $[\ ,\ ]$ on $\ggl$ 
integrates then to a Poisson-Lie structure $\Gamma^r$ on $G_r$.

One may use the right-invariant forms on $G_r$,  $\Lambda^r(X)
=\tr[(dg_+g_+^{-1}-dg_-g_-^{-1})X]$ for $X\in\ggl$, in order 
to compute the components of $\Gamma^r$. They take the form
\begin{eqnarray}\label{grps}
\gamma_{(\gp,\gm)}^r(X,Y)=tr[ X(Ad_{\gp}-Ad_{\gm})
(r_+ Ad^{-1}_{\gm}- r_- Ad^{-1}_{\gp}) Y ]. 
\end{eqnarray}

Another point from the Poisson-Lie group theory that we shall need
is the description of the Heisenberg double. In our case we shall take
$D=G\times G$ with the subgroups $G_r$ and $G_d=\{(g,g)\,|\, g\in G\}$ 
corresponding to $G^*$ and $G$. Since the intersection of the
Lie algebras $\ggl_r\cap\ggl_d$ is trivial, the intersection 
$G_r\cap G_d$ is composed of elements $(g_0,g_0)$ where $g_0$ belongs
to a discrete subgroup $G_0\subset G$. For a later use, let us note 
that if $(r_+X,r_-X)\in\ggl_r$ then $(Ad_{g_0}r_+X,Ad_{g_0}r_-X)$
also belongs to $\ggl_r$ and hence is equal to $(r_+Ad_{g_0}X,
r_-Ad_{g_0}X)$. This implies that
\begin{eqnarray}\label{com}
r_\pm Ad_{g_0}=Ad_{g_0}r_\pm.
\end{eqnarray}

We shall be interested in the description 
of the Poisson structure of $D$ on its main symplectic leaf $D_0=G_dG_r
\cap G_rG_d$ which contains a neighborhood of the unit of $D$. In $D_0$ 
the Poisson bracket is non degenerate and therefore it may be inverted 
to produce the corresponding symplectic structure $\Omega$. 
If we take $(h\gp,h\gm)=(\tilde g_{_+}\tilde h,
\tilde g_{_-}\tilde h)\in D_0$, then
\begin{eqnarray}\label{Omega1}
\Omega((h\gp,h\gm))&=&\tr[dhh^{-1}(d\tilde g_{_+}\tilde g_{_+}^{-1}-
d\tilde g_{_-}\tilde g_{_-}^{-1})\ \cr&&\hskip 5mm 
+\tilde h^{-1}d\tilde h ( \gpinv d\gp-{\gminv}
d{\gm})]
\end{eqnarray}
Note that the above form is indeed well defined on $D_0$ and that it 
is invariant under the left and right diagonal actions of $G_{0}$.

Consider the map 
\begin{eqnarray}\label{pi}
\pi: G_r\ni(\gp,\gm)\mapsto \gm\gp^{-1}\in G.
\end{eqnarray}
$\pi(g_+,g_-)=\pi(\tilde g_+,\tilde g_-)$ if and only
if $\tilde g_\pm=g_\pm g_0$ for $g_0\in G_0$. It is easy to see
that $\pi$ is a submersion so that $\pi(G_r)$ is an open connected 
subset of $G_r$ containing the unit. In fact, $\pi:G_r\rightarrow\pi(G_r)$ 
is a $G_{0}$ principal bundle, hence a discrete covering of 
$\pi(G_r)\cong G_r/G_0$. The symplectic leaves of the Poisson structure 
$\Gamma^r$ on $G_r$ are connected components of the pre-images by $\pi$
of conjugacy classes in $G$ \cite{S-T-S}. The Poisson structure 
$\Gamma^r$ projects by $\pi$ to a Poisson structure on $\pi(G_r)$ whose
contraction with $\Lambda(X)\wedge\Lambda(Y)$ is
\begin{eqnarray}\label{pspr}
\gamma^r_g(X,Y)=\tr[X(r_--Ad_gr_+)(1-Ad_g^{-1})Y].
\end{eqnarray}
The symplectic leaves of the projected structure are the connected
components of intersection of $\pi(G_r)$ with the conjugacy classes in $G$.
In fact, formula (\ref{pspr}) defines a Poisson structure on the whole 
group manifold $G$. This is a different Poisson structure on $G$
than that given by (\ref{Gamma}). In particular, it does not make 
$G$ with its standard multiplication a Poisson-Lie group.

In the connection between topological Poisson-sigma model with 
the dual group $G^*$
as the target and the gauged group $G$ WZW model, the operator $r$ plays 
the crucial role. It determines, as we have seen, the Poisson structure
of the target in the former and in the latter it may be used to write 
the closed form $\chi={1\over 3}\tr[(g^{-1}dg)^{\wedge3}]$ on $G$ 
entering the Wess-Zumino term of the action as a differential of a 
2-form on the open subset $\pi(G_r)$ of $G$. More exactly,
the pullback of $\chi$ by $\pi$ has the form
\begin{eqnarray}\label{pichi}
\pi^*\chi&=&{1\over 3}\tr[(\gminv d\gm -\gpinv d\gp)^{\wedge3}]\cr
           &=&d\ \tr[\gminv d\gm \wedge\gpinv d\gp]\cr
&&+{1\over 3}\tr[(\gminv d\gm)^{\wedge3}]
  -{1\over 3}\tr[(\gpinv d\gp)^{\wedge3}].
\end{eqnarray}
Let us show that the last two terms cancel each other.
Define 
$\mu=\gpinv d\gp -\gminv d\gm\in\Lambda^1(G_r)\otimes{\ggl}$.
One has $r_{\pm}\mu=g_{_\pm}^{-1}dg_{_\pm}$ and 
\begin{eqnarray}
\tr[(\gpinv d\gp)^{\wedge3}]&=&\tr[(r_+\mu)^{\wedge3}]\cr
                    &=&{1\over 2}\tr[r_+(\mu\wedge r\mu+r\mu\wedge\mu)
                           \wedge r_+\mu]\cr
                    &=&-{1\over 2}\tr[(\mu\wedge r\mu + r\mu\wedge\mu)
                             \wedge r_-r_+\mu]
\end{eqnarray}
where we have used (\ref{homo}) and the antisymmetry of $r$
with respect to the bilinear form $\tr$.
The same result is obtained for $\tr[(\gminv d\gm)^{\wedge3}]$
(of course $r_+r_-=r_-r_+$). Coming back to (\ref{pichi})
we infer that $\chi= d\rho$ on $\pi(G_r)$ where $\rho$ is a 2-form
on $\pi(G_r)$ such that
 \begin{eqnarray}\label{rho}
\pi^*\rho=\tr[\gminv d\gm \wedge\gpinv d\gp]\,.
\end{eqnarray}

A straightforward computation leads to an equivalent expression 
for the symplectic form $\Omega$ of (\ref{Omega1}) on the leaf
$D_0$ in the Heisenberg double:
\begin{eqnarray}\label{Omega2}
\Omega&=&\tr[(g^{-1}dg+dgg^{-1}+g^{-1}h^{-1}dhg)h^{-1}dh]\cr&&
\hskip -2mm+\,\rho(g)-\rho(hgh^{-1}).
\end{eqnarray}               
The right hand side may be viewed as the 2-form on
the space of pairs $(h,g)$ such that both $g=g_-g_+^{-1}$ and $hgh^{-1}
=\tilde g_-{\tilde g}_+^{-1}$ are in $\pi(G_r)$. We shall denote it then
as $\Omega(h,g)$. It is in this form that the symplectic form of the
Heisenberg double will appear below.

\nsection{Restricted $\,G/G\,$ coset theory}

The $G/G$ coset theory is the gauged version of the WZW model \cite{WZW}
with the group $G$ as the target. The fields of this model on a 
two-dimensional oriented surface $\Sigma$
equipped with a conformal or a pseudo-conformal structure are 
$g:\Sigma\rightarrow G$ and a gauge field $A$, a $\ggl$-valued 1-form 
on $\Sigma$. The action functional of the model
has the form
\begin{eqnarray}\label{swzw}
S_{WZW}(g,A)&=& {1\over4\pi}\int\limits_\Sigma\tr[(g^{-1}\parl g)(g^{-1}
\parr g)]dx^l\wedge dx^r+S_{WZ}(g)\cr\cr
&&\hskip -3mm+{1\over 2\pi}\int\limits_\Sigma\tr[\parl g g^{-1}\AR - \AL 
g^{-1}\parr g]dx^l\wedge dx^r\cr\cr 
&&\hskip -3mm+{1\over 2\pi}\int\limits_\Sigma\tr[\AL\AR  - g\AL g^{-1}\AR]
dx^l\wedge dx^r
\end{eqnarray}
where  $x^l=z$, $x^r=\bar z$ are the complex variables in the Euclidean 
signature and the light-cone ones $x^l=x+t$, $x^r=x-t$ in the Minkowski 
signature. The derivatives $\parl={\partial\over\partial 
x^l}$, $\parr={\partial\over\partial x^r}$ and $A=A_ldx^l+A_rdx^r$. 
The Wess-Zumino term $S_{WZ}(g)$ is often written as
${1\over 4\pi}\int\limits_\Sigma g^*\,d^{-1}\chi$ where 
$d^{-1}\chi$ stands for a 2-form whose differential is equal to the 
3-form $\chi$ on $G$. Since such 2-forms do not exist globally, some 
choices are needed. On closed surfaces, one may replace 
$\int\limits_\Sigma g^*d^{-1}\chi$ by $\int\limits_B{\tilde g}^*\chi$ 
where $B$ is an oriented 3-manifold with $\partial B=\Sigma$ and 
$\tilde g:B\rightarrow G$ extends $g$. This gives a well defined 
amplitude $\ee^{ikS_{WZW}}$ for integer $k$. When $\Sigma$ has 
a boundary, the definition of the amplitude is more problematic and, 
in general, it makes sense only as an element of a product of line bundles
over the loop group $LG$ \cite{Gaw0}. To extract a numerical value of such
an amplitude one has to use sections of the line bundle that are not
globally defined. 

What we shall do here, is to restrict the values of the field $g$
of the gauged WZW model to the subset $\pi(G_r)\subset G$ and to define
\begin{eqnarray}\label{wz}
S_{WZ}(g)={1\over 4\pi}\int\limits_\Sigma g^*\rho
\end{eqnarray}
where $\rho$ is given by (\ref{rho}). In the geometric language, 
this corresponds to a particular choice of a section in the (trivial) 
restriction of the line bundle over $LG$ to the loop 
space $L\pi(G_r)$. We shall call the resulting field theory the 
restricted $\,G/G\,$ coset model. 

The model defined above has simple transformation properties under
the gauge transformations of fields
\begin{eqnarray}\label{gauge}
{}^h\hspace{-0.05cm}g=h gh^{-1}\,\qquad 
{}^h\hspace{-0.08cm}A=h Ah^{-1}+h dh^{-1}
\end{eqnarray}
for $h:\Sigma\rightarrow G$ such that ${}^h\hspace{-0.05cm}g$ takes 
also values in $\pi(G_r)$ (this is always 
accomplished for $h$ sufficiently close to unity). The action transforms
according to 
\begin{eqnarray}\label{Sgtr}
S_{WZW}({}^h\hspace{-0.05cm}g,{}^h\hspace{-0.08cm}A)=S_{WZW}
(g,A)+\int_\Sigma \Omega(h,g)
\end{eqnarray}
where $\Omega(h,g)$ is the closed 2-form given by (\ref{Omega2}).
In particular, it follows that the action is invariant under infinitesimal 
gauge transformations that vanish on the boundary.

The equations of motion  of the model $\delta S_{WZW}=0$ for field
variations vanishing at the boundary are
\begin{eqnarray}
&D_l(g^{-1}D_rg)dx^l\wedge dx^r +F(A)=0&\alabel{eom}{a}\cr
&g^{-1}D_rg=0\,,\qquad gD_lg^{-1}\,=0&\aeqno{b}\cr
\end{eqnarray}
\vskip -0.37cm
\noindent where $D$ stands for the covariant derivative
$D_{l,r}=\partial_{l,r} + [A_{l,r},\,\cdot\,]$ 
and $F(A)=dA+A\wedge A$ is the field strength of $A$.

Although (\ref{eom},a) is a second order differential equation we can 
write a system of first order equations equivalent to (\ref{eom}),
by simply taking
\begin{eqnarray}
&&F(A)=0,\alabel{curv}{a}\cr
&&g^{-1}Dg=0.\aeqno{b}\cr
\end{eqnarray}
\vskip -0.37cm
\noindent The situation we find is analogous to the following simple
mechanical system. Starting from Lagrangian
$$L={1\over2} \dot x^2+ y \dot x$$
the equations of motion are
$$ \ddot x + \dot y = 0\qquad \dot x=0$$
or equivalently
$$ \dot y=0\qquad \dot x=0\,.$$
As we have first order equations of motion it 
should be possible to get the system from a first order Lagrangian.
We can achieve this goal by adding to the Lagrangian terms quadratic
in the equations of motion which do not change
the dynamics. The phase space structure of the theory
remains then unchanged. In the Hamiltonian language, all we do is to 
trade secondary constraints for primary ones, which does not make
any difference in the canonical analysis of the theory.
In our simple example, the new Lagrangian 
$$L'=y \dot x$$ 
is equivalent to the previous one. Note, on the other
hand, that the new action is topological in the sense
that it is invariant under diffeomorphism of time.

In the following we will proceed in a similar
way for the restricted $G/G$ coset model, using its first
order equations of motion to eliminate from the action the piece
quadratic in derivatives. We shall identify the result
with a topological Poisson sigma model with group $G_r$ equipped
with the Poisson structure $\Gamma^r$ as the target.
We shall call the latter theory the topological Poisson-Lie 
sigma model.

\nsection{Topological Poisson-Lie sigma model}

In general, the Poisson sigma model \cite{SS}\cite{cafe}
is a two-dimensional sigma model with the target manifold $\CX$
equipped with a Poisson structure $\Gamma$. The fields of the model  
are $X:\Sigma\rightarrow\CX$ and a 1-form $\psi$ on $\Sigma$ with 
values in the pullback by $X$ of the cotangent bundle of $\CX$. 
The action functional has the form
\begin{eqnarray}\label{spsg}
S_{P\sigma}(X,\psi)=-{1\over2\pi}\int\limits_\Sigma\Big[\langle dX,\wedge 
\psi\rangle+{1\over2}\langle\Gamma\circ X,\psi\wedge\psi\rangle\Big]
\end{eqnarray}
where $\langle\ \hspace{0.03cm},\ \rangle$ denotes the pairing between 
the tangent and the cotangent vectors to $\CX$. 

We shall be interested in the Poisson-Lie case where $\CX=G_r$ 
and $\Gamma=\Gamma^r$. In this case, $X=(g_+,g_-)$. The tangent bundle 
to $G_r$ may be identified with the trivial bundle $G_r\times\ggl_r$ 
by the right translations. With this identification, $dX=(dg_+g_+^{-1},
dg_-g_-^{-1})$. Similarly, the cotangent bundle to $G_r$
may be identified by the right translations with the trivial bundle 
$G_r\times\ggl$ and, using this identification, field $\psi$ becomes 
a $\ggl$-valued 1-form that we shall denote by $A$.
Recalling the pairing (\ref{pair}) between $\ggl_r$ and $\ggl$
and the formula (\ref{grps}) for the components of $\Gamma^r$, we infer 
that
\begin{eqnarray}\label{sps}
S_{P\sigma}(g_+,g_-,A)&=&-{1\over 2\pi}\int\limits_\Sigma
\tr[(d\gp\gpinv -d\gm\gminv)\wedge A
\cr\cr&&\hspace{-2cm}
+{1\over2}
A\wedge(\Ad_{\gp}-\Ad_{\gm})(r_+\Ad_{\gm}^{-1} -r_-\Ad_{\gp}^{-1})A].
\end{eqnarray}

The Poisson-Lie sigma model with $G_r$ target and fields
$(g_+,g_-)$ and $A$ is closely related to the gauged WZW model with 
the target $\pi(G_r)$ and fields $g=g_-g_+^{-1}$ and $A$ that we considered
in the previous section. The action of the latter is given by equation 
(\ref{swzw}) with the Wess-Zumino term defined by (\ref{wz}). More
explicitly,
\begin{eqnarray}\label{swzwpm}
S_{WZW}(g_-g_+^{-1},A)&=& {1\over4\pi}\int\limits_\Sigma
\tr[(\jlp)(\jrp)+(\jlm)(\jrm)
\cr&&\hskip 15mm -2(\jlp)(\jrm)]
dx^l\wedge dx^r\cr 
&&\hskip -3mm -{1\over 2\pi}\int\limits_\Sigma\tr[\AL g^{-1}\parr g+
(g\parl g^{-1})\AR 
 \cr&&\hskip 16mm +g\AL g^{-1}\AR -\AL\AR]dx^l\wedge dx^r.
\end{eqnarray}
The equations of motion are (\ref{curv},a) and (\ref{curv},b).
The latter reads in terms of $(\gp,\gm)$ as
\begin{eqnarray}\label{eompm}
P_{l ,\pm}&\equiv&g_\pm^{-1}\parl g_{\pm}+
r_\pm (\Ad_{\gp}^{-1}-\Ad_{\gm}^{-1})\AL = 0,
\cr
P_{r ,\pm}&\equiv&g_\pm^{-1}\parr g_{\pm}+ 
r_\pm (\Ad_{\gp}^{-1}-\Ad_{\gm}^{-1})\AR = 0.
\end{eqnarray}
After a straightforward calculation one obtains the identity:
\begin{eqnarray}
S_{WZW}(g_-g_+^{-1},A)&=& {1\over4\pi}\int\limits_\Sigma\tr(
P_{l,+}P_{r,+}+P_{l,-}P_{r,-}-2 P_{l,+}P_{r,-})dx^l\wedge dx^r\cr
&&\hspace{4cm}+\ S_{P\sigma}
(g_+,g_-,A)
\end{eqnarray}
where $S_{P\sigma}$ is the action (\ref{sps}) of the Poisson-Lie
sigma model. 

Note that the difference between $S_{WZW}$ and $S_{P\sigma}$ is
quadratic in the equations of motion for $S_{WZW}$. This implies 
that neither the classical solutions nor the phase space structure 
of both models (based on first functional derivatives of the action 
evaluated on-shell) differ, as long as the boundary conditions 
coincide. The latter will be chosen to require in both models that 
$A$ vanishes on vectors tangent to boundary, with no conditions 
on $g$ nor $(g_+,g_-)$, see below. The relation between the gauged 
WZW model and the $G_r$ Poisson sigma model was first established 
in \cite{AlSchStr} in the Hamiltonian formalism. The above discussion 
provides a covariant version of that result valid in an arbitrary 
space-time topology. 

A warning is in order here: the equivalence of the models
is restricted for the moment to fields $g$ of the $G/G$ theory
which are of the form $g_-g_+^{-1}$ for $(g_+,g_-):\Sigma\rightarrow G_r$. 
The general $\pi(G_r)$-valued fields $g$ have such form only locally, 
with local representatives possibly differing by the right action 
of elements $g_0\in G_0$. To extend the relation to the general 
case, let us replace the gauge field $A$ by another $\ggl$-valued 
1-form
\begin{eqnarray}\label{etaA}
\eta=\Ad_{g_-} (r_-\Ad_{g_+}^{-1}-r_+\Ad_{g_-}^{-1})A,
\end{eqnarray}
Given the $\pi(G_r)$-valued field $g$, 1-form $\eta$
is unambiguously defined, as follows from (\ref{com}). Conversely,
the gauge field $A$ may be recovered from $\eta$\,:
\begin{eqnarray}\label{Aeta}
A=(r_-\Ad_{g}^{-1}-r_+)\eta.  
\end{eqnarray}
Actually (\ref{etaA}) and (\ref{Aeta}) solve the equation
\begin{eqnarray}
\tr(dgg^{-1}\wedge\eta)=\tr[(d\gp\gpinv -d\gm\gminv)\wedge A].
\end{eqnarray}
Now, in terms of $g$ and $\eta$,
\begin{eqnarray}\label{spseta}
S_{P\sigma}&=&-{1\over 2\pi}\int_\Sigma\tr[(dgg^{-1}\wedge\eta)
+{1\over2}
\eta\wedge(r_--\Ad_gr_+)(1-\Ad_g^{-1})\eta].\hspace{0.3cm}   
\end{eqnarray}
We recognize the action of the Poisson sigma model with the target 
$\pi(G_r)\cong G_r/G_0$ equipped with the Poisson structure (\ref{pspr}) 
projected from $G_r$. The relations (\ref{etaA}) and (\ref{Aeta}) define 
now a one-to-one correspondence between the classical solution $(g,A)$ 
and of the gauged WZW theory and the solution $(g,\eta)$ of the Poisson 
sigma model with the action (\ref{spseta}), both theories with 
$\pi(G_r)$ target.

\nsection{Canonical structure}

Having established the connection between the restricted $G/G$ coset
theory and the orbifold version of the Poisson-Lie sigma model with 
$G_r$ target, we shall study the phase space structure of the 
theory on space-times $\Sigma=\NR\times(\NR/2\pi\NZ)$ (closed geometry) 
or $\Sigma=\NR\times[0,\pi]$ (open geometry), with the coordinates 
$(t,x)$ and Minkowski signature. We shall use the first order 
formalism, specially convenient for extracting the symplectic 
structure of the phase space, see \cite{Gaw}. 

In this formalism, the basic object is a 2-form $\alpha$ 
on a space $\CP$ which one may take as a bundle over the space-time 
$\Sigma$. (In general space-time dimension 
$d$, $\alpha$ is a $d$-form). Fields $\Phi$ are sections of $\CP$ 
and the first order action is given by
\begin{eqnarray}\label{1oa}
S(\Phi)=\int_{\Sigma}\Phi^*\alpha
\end{eqnarray}
In the open geometry, fields $\Phi$ may be required 
to satisfy boundary conditions restricting their values on 
$\partial\Sigma$ to a subbundle $\CP_\partial$ of $\,\CP|_{\partial\Sigma}$.
A term $\int_{\partial\Sigma}\Phi^*\beta$ where $\beta$ is a 1-form
on $\CP_\partial$ ($(d-1)$-form in general) may be added to the action.
The variational equations $\delta S(\Phi)=0$ take the geometric form  
\begin{eqnarray}
&&\hbox to 4.4 cm{$\Phi^*(\iota(\delta\Phi)d\alpha)=0$\hfill}
{\rm on}\quad\Sigma\alabel{eq1}{a}
\cr\cr
&&\hbox to 4.4 cm{$\Phi^*(\iota(\delta\Phi)(\alpha+d\beta)=0$\hfill}
{\rm along}\quad\partial\Sigma\aeqno{b}\cr
\end{eqnarray}
\vskip -0.37cm
\noindent for any vector fields $\delta\Phi$ giving the infinitesimal 
variation of $\Phi$ tangent on $\partial\Sigma$ to the subbundle
$\CP_\partial$. Let us denote by $\tilde P$ the space of the solutions
of these equations defined on $\Sigma$ (and subject to the boundary 
condition). 

Let now $\Sigma_{12}$ be a a chunk of $\Sigma$ between two  
(maximal) curves ($(d-1)$-dimensional surfaces, in general) $\CC_1$ and 
$\CC_2$ transverse to the time axis. Let $S_{12}(\Phi)$ be defined
by restricting the integration in (\ref{1oa}) to $\Sigma_{12}$
(and to $\partial\Sigma\cap\Sigma_{12}$ in the $\beta$-term).
If $\Phi$ belongs to the space $\tilde P$ of solutions and the vector
field $\delta\Phi$ gives a variation tangent to $\tilde P$ then, as 
a straightforward calculation shows,
\begin{eqnarray}
\delta S_{12}(\Phi)=\Xi_{\CC_2}(\delta\Phi)-\Xi_{\CC_1}(\delta\Phi)
\end{eqnarray}
where $\Xi_\CC$ is a 1-form on the space of solutions defined by
\begin{eqnarray}
\Xi_\CC(\delta\Phi)=\int\limits_\CC\Phi^*(\iota(\delta\Phi)\alpha)
-\int\limits_{\partial\CC}\Phi^*(\iota(\delta\Phi)\beta).
\end{eqnarray}
Clearly $\tilde\Omega=d\Xi$ is a closed 2-form on $\tilde P$
that does not depend on the choice of $\CC$. The explicit
formula for $\tilde\Omega$ reads \cite{GTT}
\begin{eqnarray}\label{Omegagen}
\widetilde\Omega(\delta_1\Phi,\delta_2\Phi)&=&\int_\CC\Phi^*(
\iota(\delta_2\Phi)\iota(\delta_1\Phi)d\alpha))\cr
&-&\int_{\partial\CC}\Phi^*(
\iota(\delta_2\Phi)\iota(\delta_1\Phi)(\alpha+d\beta))\,.
\end{eqnarray}
The 2-form $\tilde\Omega$ does not have to be non-degenerate,
but, since it is closed, its degeneration distribution is involutive. 
By definition, the phase space $P$ of the field theory is the space 
of leaves of this distribution. $\widetilde\Omega$ descends to 
a symplectic form on $P$ (more exactly, on its non-singular part).

In our case, we could use either the Poisson-Lie formulation of the 
theory whose action is already first order in field derivatives or the 
coset one. The latter requires introduction of variables $\xi_{l,r}$ 
representing the light-cone derivatives of field $g$ to obtain its 
first order formulation. As stressed above, both approaches 
lead to the same phase space. 

\subsection{Poisson sigma models}

Let us start by few comments about
the canonical structure of the topological Poisson sigma models 
with a general target $\CX$, an interesting issue in its own.

For a Poisson sigma model we take 
$\CP=\Sigma\times T^*\CX\oplus T^*\CX$
with local coordinates $(x^l,x^r,X^a,\psi_{al},\psi_{ar})$ and
\begin{eqnarray}\label{alo}
\alpha=-{1\over 2\pi}\Big[dX^a\wedge(\psi_{al}dx^l+\psi_{ar}dx^r)
+\Gamma^{ab}(X)\psi_{al}\psi_{br}dx^l\wedge dx^r\Big]
\end{eqnarray}
so that the action (\ref{1oa}) with $\beta=0$ reproduces the action
(\ref{spsg}) with $\psi=\psi_ldx^l+\psi_rdx^r=\psi_tdt+\psi_xdx$. 
In the bulk, the equations 
of motion are \cite{SS}
\begin{eqnarray}
&\partial_\mu X^a=\Gamma^{ab}(X)\psi_{b\mu}\,,\alabel{eqps}{a}\cr\cr
&\partial_\mu\psi_{a\nu}-\partial_\nu\psi_{a\mu}=\partial_a\Gamma^{bc}(X)
\psi_{b\mu}\psi_{c\nu}.&\aeqno{b}\cr
\end{eqnarray}
\vskip -0.37cm
\noindent In the open geometry, one has, additionally, the boundary 
equation 
\begin{eqnarray}\label{bde}
\delta X^a\psi_{at}(t,0)=0=\delta X^a\psi_{at}(t,\pi)\,.
\end{eqnarray}
It will be solved by imposing the condition $\psi_t=0$ and keeping 
$X$ free on the boundary. On more general surfaces $\Sigma$ with 
boundary, we shall require that $\psi$ vanishes on vectors tangent 
to $\partial\Sigma$. Such boundary conditions were used in \cite{cafe} 
in connection with the Kontsevich deformation quantization \cite{Konts}. 

Note that $X$ evolves within fixed symplectic leaves
and that it is constant on the boundary. The 2-form $\widetilde\Omega$ 
on the space of solutions is
\begin{eqnarray}\label{ops1}
\widetilde\Omega={1\over2\pi}\int\limits_\CC dX^a\wedge d\psi_{ax}\, dx
\end{eqnarray}
with the integral over any maximal curve of constant $t$. The form
$\widetilde\Omega$ has a large degeneration kernel given by the
vector fields
\begin{eqnarray}
\delta X^a=\Gamma^{ab}(X)\epsilon_{b}\qquad
\delta\psi_{a\mu}=\partial_\mu\epsilon_a-\partial_a\Gamma^{bc}
\psi_{b\mu}\epsilon_{c}
\label{deg}
\end{eqnarray}
with $\epsilon_a(t,x)$ vanishing on the boundary in the
open geometry. The latter should be viewed as corresponding 
to the infinitesimal gauge transformations in the Poisson 
sigma model context \cite{SS}. The phase space of the theory $P$ is 
the space of the classical solutions modulo the degeneration 
induced by the vector fields (\ref{deg}). At non-singular points,
the tangent space to $P$ may be identified with the space
of vector fields $(\delta X,\delta\psi_x)$ at fixed $t$ such that
\begin{eqnarray}\label{D1}
D_1(\delta X,\delta\psi_{x})\equiv
\partial_x\delta X^a-\Gamma^{ab}(X)\delta\psi_{bx}-\partial_c
\Gamma^{ab}(X)\psi_{bx}\delta X^c=0
\end{eqnarray}
modulo
\begin{eqnarray}\label{D0}
(\delta X,\delta\psi_{x})=(\Gamma^{ab}(X)\epsilon_b,\,\partial_x
\epsilon_a-\partial_a\Gamma^{bc}(X)\psi_{bx}\epsilon_c)\equiv D_0\,
\epsilon
\end{eqnarray}
with $\epsilon(t,0)=\epsilon(t,\pi)=0$ in the open geometry.
Operators $(D_0,D_1)$ form an elliptic complex. We infer that the phase 
space $P$ is finite-dimensional in non-singular points and has
dimension equal to the index of $(D_0,D_1)$. 

It is easy to compute the latter at special solution 
$X=X_0=const.$, $\psi=0$ of the classical
equations. Let $(\delta X,\delta\psi_x)$ be a corresponding
solution of (\ref{D1}) in the open geometry. By subtracting the pair 
(\ref{D0}) for a unique $\epsilon$ vanishing on the boundary, 
$(\delta X,\delta\psi_x)$ may be brought 
to a linear form $(\delta'X,\delta'\psi_x)$ with
\begin{eqnarray}
\delta'X^a(x)=x\Gamma^{ab}(X_0)\delta'\psi_{bx}+\delta X^a(0),
\qquad\delta'\psi_x={1\over\pi}\int\limits_0^\pi\delta\psi_x\,dx
\end{eqnarray}
otherwise unconstrained. It follows that the virtual dimension
of $P$ is equal to $2\,dim(\CX)$ in this case.
As explained in \cite{cafe2}, see also \cite{cafe3}, 
symplectic space $P$ may be identified for the open geometry 
with the Poisson groupoid of $\CX$ (again modulo singularities). 
In particular, a point in $P$ contains the information about 
the homotopy class $[X]$ of paths in a fixed symplectic leaf 
of $\CX$ obtained by restricting the solution $X$ to any spatial 
hypercurve. Such paths run between the ($t$-independent) points 
$X(t,0)$ and $X(t,\pi)$. The homotopy classes $[X]$ may be 
composed if the end of one agrees with the beginning of the 
other and this composition is consistent with the groupoid 
structure in $P$. We refer for the details to \cite{cafe2}.

Similarly, in the closed geometry, one may achieve that
\begin{eqnarray}
\delta' {X}^a(x)=\delta X^a(0)-\Gamma^{ab}(X_0)\epsilon_b(0),
\qquad\delta'\psi_x={1\over2\pi}\int\limits_0^{2\pi}\delta\psi_x\,dx
\end{eqnarray}
but now with the constraint $\Gamma^{ab}(X_0)\delta'\psi_{bx}
=0$ that follows from (\ref{D1}) and the remaining gauge
transformations by constant $\epsilon$. The virtual dimension of $P$ 
is then equal to $2(dim(\CX)-dim(\CL))$, where $\CL$ 
is the symplectic leaf of $\CX$ containing $X_0$. The non-singular 
part of $P$ is a symplectic manifold of dimension equal to twice 
the transverse dimension of the foliation of $\CX$ by the symplectic 
leaves. A point in $P$ contains the information about the homotopy 
class $[X]$ of loops in a symplectic leaf. In particular, there 
is a natural projection from $P$ to the space of leaves and 
a natural injection of the latter into $P$ obtained by assigning 
the constant solutions $X=X_0$, $\psi=0$  to points $X_0\in\CX$. Away 
from the singularities, this injection is a local embedding onto 
a Lagrangian submanifold of $P$. Altogether, one obtains an interesting
structure that certainly deserves further study and a name 
in the general context of Poisson manifolds. We propose to call
the closed-geometry phase space $P$ the transversal of the Poisson
manifold $\CX$.

The general discussion applies, of course, to the case of 
Poisson-Lie model (\ref{sps}) with target $G_r$ or
to its orbifold (\ref{spseta}) with target $G_r/G_0\cong\pi(G_r)$
In particular, the latter has the same phase space with
the same symplectic structure as the restricted $G/G$ coset model.
In the $G/G$ model, the degeneracy directions in the space of classical
solutions are given by the standard gauge transformations and
the quotient is somewhat easier to describe explicitly, as we shall
do now.

\subsection{$G/G$ \,coset theory}

In the first order formulation of the $G/G$ theory we consider 
the space $\CP=\Sigma\times \pi(G_r)\times{\gl g}^4$ of points 
$({\bf x},g,\xi_l,\xi_r,A_l,A_r)$, where ${\bf x}=(x^l,x^r)$, 
the 2-form 
\begin{eqnarray}
\alpha&=&{1\over 4\pi} \tr[-\xi_l\xi_r dx^l\wedge dx^r
-\xi_l\, g^{-1}dg\wedge dx^l +\xi_r\, g^{-1}dg\wedge dx^r]
\cr\cr&&\hskip -3mm+{1\over4\pi}\rho(g)\,+\, {1\over 2\pi}
\tr[A_l dgg^{-1}\wedge dx^l + A_r g^{-1}dg\wedge dx^r  
\cr&&\hskip 26mm
+ A_lA_r dx^l\wedge dx^r-gA_lg^{-1}A_r dx^l\wedge dx^r]
\end{eqnarray}
with vanishing 1-form $\beta$. For a field configuration 
$$\Phi({\bf x})=(g({\bf x}),\xi_l({\bf x}),
\xi_r({\bf x}), A_l({\bf x}), A_r({\bf x}))\,,$$
the first order action (\ref{1oa}) coincides with the second-order 
one of (\ref{swzw}) if 
\begin{eqnarray}\label{eomfo}
\xi_l=g^{-1}\partial_l g\quad\mbox{and}\quad\xi_r=g^{-1}\partial_r g\,.
\end{eqnarray}
In the first order approach we do not impose those equations
but extract them from the extremality condition $\delta S(\Phi)=0$
which, in our case, imply relations (\ref{eomfo})
together with the old equations (\ref{curv}) and, in the open
geometry, impose also the boundary equation 
\begin{eqnarray}\label{boucon}
\tr[(\delta g_+g_+^{-1}-\delta g_-g_-^{-1})A_t]=0\qquad {\rm on}\quad 
\partial \Sigma
\end{eqnarray}
for any $\delta g$ compatible with the boundary conditions.
We solve (\ref{boucon}) by taking $A_t=A_l-A_r=0$ on the boundary 
of  $\Sigma$ and $g$ free. These boundary conditions correspond to
the ones in the the Poisson sigma models considered above.
\vskip 0.4cm

\noindent{\bf{Closed geometry}}
\vskip 0.3cm

In order to solve the classical equations (\ref{curv})
on the cylinder, we introduce the space $\CG$ of maps $\hat h:\NR^2
\rightarrow G$ such that $\hat h(t,x+2\pi)=\hat h(t,x)h_1^{-1}$ 
for some $h_1\in G$ and all $(t,x)$. 
A general solution of (\ref{curv},a) has then the form 
\begin{eqnarray}\label{sA}
A=\hat hd\hat h^{-1}, \qquad \hat h\in \CG.
\end{eqnarray}
The solutions of (\ref{curv},b) are in turn given by
\begin{eqnarray}\label{sg}
g=\hat h\hat g{\hat h}^{-1},\qquad\hat g\in G 
\end{eqnarray}
with constant $\hat g$, provided that $g(t,x)$ takes values in 
$\pi(G_r)$. In particular, it follows that $\hat g$ commutes
with $h_1$. All pairs $(\hat h\gamma^{-1},\gamma\hat g\gamma^{-1})$
for constant $\gamma\in G$ correspond to the same solution
of equations (\ref{curv}) and this is the only ambiguity. 
We may identify then the space of solutions with a subspace of 
$(\CG\times G)/G$.

The closed 2-form (\ref{ops1}) is easy to compute
by plugging in the general solution (\ref{sA}), (\ref{sg}) 
in terms of $\,\hat h(t,x)$ and $\hat g$ to the formula
(\ref{Omegagen}). One obtains:
\begin{eqnarray}\label{Omega0}
\widetilde\Omega(\hat h,\hat g)=
{1\over 4\pi}\tr[ h_1^{-1}dh_1\wedge(\hat g^{-1} d\hat g 
+ d\hat g \hat g^{-1} + \hat g^{-1}h_1^{-1}dh_1\hat g)].
\end{eqnarray}
It is easy to check that the form (\ref{Omega0}) projects to the 
orbit space $(\CG\times G)/G$. In fact, it depends only on
the orbits under simultaneous conjugations of the commuting pairs 
$(h_1,\hat g)$ of elements in $G$ . It follows that 
$\widetilde\Omega$ has a huge kernel composed of the vectors 
$(\delta\hat h,\delta\hat g)$ such that $(\delta\hat h{\hat h}^{-1})
(t,x)$ is periodic in $x$ and otherwise arbitrary and 
that $\hat g^{-1}\delta\hat g=0$. The corresponding flows generate 
the group of gauge transformations $h:\Sigma\rightarrow G$ that act 
by left multiplication on $\hat h$. It is the same local gauge symmetry 
that was already identified in previous Section in the Lagrangian 
formalism. 

We may describe now precisely the reduced phase space $P$, i.e.
the transversal of the Poisson manifold $\,\pi(G_r)\cong G_r/G_0$. 
It is composed of the $G$-orbits of pairs $([\hat h],\hat g)$ 
where $\hat g\in G$ and $[\hat h]$ is the homotopy class of maps 
$\hat h:\NR\rightarrow G$ such $\hat h\hat g\hat h^{-1}$ takes
values in $\pi(G_r)$ and that $\hat h(x+2\pi)=\hat h(x) h_1^{-1}$ 
for $h_1\in G$ commuting with $\hat g$ (the homotopies leave $h_1$ 
unchanged). The $G$-orbits are composed 
of pairs $([\hat h\gamma^{-1}],\gamma\hat g\gamma^{-1})$, $\gamma\in G$. 
The phase space $P$ covers the space of conjugacy classes
of commuting pairs $(h_1,\hat g)$.

Generic commuting pairs may be brought by conjugation to the form 
$(\ee^{2\pi\mu}, \ee^{2\pi\nu})$ with $\mu$ and $\nu$ in
the Cartan subalgebra of $\ggl$. In this parametrization, 
the expression for the symplectic form simplifies to
\begin{eqnarray}\label{Omega11}
\widetilde\Omega(\hat h,\hat g)=2\pi\,\tr[d\mu d\nu].
\end{eqnarray}
Note that at non-singular points $P$ has dimension equal to twice 
the rank of $G$. It is equal to twice the transverse dimension 
of the foliation of $\pi(G_r)$ by the symplectic leaves (the conjugacy 
classes) at generic points, in accordance with the general
discussion.
\vskip 0.4cm

\noindent{\bf{Open geometry}}
\vskip 0.3cm

On the strip, the classical equation (\ref{curv})
together with the boundary condition $A_t=0$ may be solved 
similarly. We take as $\CG$ the space of maps $\hat h:\Sigma\rightarrow 
G$ constant on every connected component of $\partial \Sigma$, i.e. 
$\hat h(t,0)=h_0$ and $\hat h(t,\pi)=h_\pi$. Then (\ref{sA})
and (\ref{sg}) still gives a general solution of (\ref{curv}),
provided that $g$ takes values in $\pi(G_r)$. All pairs 
$(\hat h\gamma^{-1},\gamma\hat g\gamma^{-1})$, $\gamma\in G$, 
correspond to the same solution. The space of solutions 
may be again identified with a subspace of $(\CG\times G)/G$. 
Of course we could fix now the $\gamma$-ambiguity by restricting 
$\hat h$ to be $1$ on one of the component of the boundary but, for 
a moment, we prefer not to make any choice. 

The 2-form (\ref{Omegagen}) is for the open geometry given by
\begin{eqnarray}\label{Omega}
\widetilde\Omega(\hat h,\hat g)&=&{1\over 4\pi}
\tr[(\hat g^{-1} d\hat g + d\hat g \hat g^{-1} + 
                  \hat g^{-1}h_\pi^{-1}dh_\pi\hat g)\wedge h_\pi^{-1}dh_\pi]-
 {1\over 4\pi}
\rho(h_\pi\hat g h_\pi^{-1})
\cr\cr&&\hspace{-1.4cm}
-{1\over 4\pi}\tr[(\hat g^{-1} d\hat g + d\hat g \hat g^{-1} + 
                   \hat g^{-1}h_0^{-1}dh_0\hat g)\wedge h_0^{-1}dh_0]+
 {1\over 4\pi}
\rho(h_0\hat g h_0^{-1})
\end{eqnarray}  
which, again, is unambiguously defined on $(\CG\times G)/G$.
Indeed, it descends to the quotient space $(G\times G\times G)/G$ 
of the $G$-orbits of triples $(h_0,h_\pi,\hat g)$. 
The kernel of $\widetilde\Omega$ is given by the vectors
$(\delta\hat h,\delta\hat g)$ such that $(\delta\hat 
h{\hat h}^{-1})(t,x)$ vanishes at $x=0$ and $x=\pi$ and that 
$g^{-1}\delta\hat g=0$. They generate the gauge transformations 
$h:\Sigma\rightarrow G$ equal to $1$ on the boundary and acting 
by left multiplication on $\tilde h$.
 
The reduced phase space $P$ of the model is composed of the $G$-orbits
of pairs $([\hat h],\hat g)$ where $[\hat h]$ is a homotopy class
of paths $\hat h:[0,\pi]\rightarrow G$ such that $\hat h\hat g\hat h^{-1}$
takes values in $\pi(G_r)$. The homotopies are supposed to be fixed
on the boundary. Space $P$ covers the space of $G$-orbits of triples
$(h_0,h_\pi,\hat g)$ such that $h_0\hat gh_0^{-1}$ and $h_\pi\hat g
h_\pi^{-1}$ lie in the same connected component of $\pi(G_r)\cap 
C_{\hat g}$ where
$C_{\hat g}$ is the conjugacy class of $\hat g$. Recall that such 
components form the symplectic leaves of the Poisson structure (\ref{pspr}) 
on $\pi(G_r)$. There are two natural maps from $P$ to $\pi(G_r)$ induced by 
\begin{eqnarray}
([\hat h],\hat g)\ \mapsto\ h_0\hat gh_0^{-1}\qquad{\rm and}
\qquad([\hat h],\hat g)\ \mapsto\ h_\pi\hat gh_\pi^{-1}
\end{eqnarray}
that map into the same symplectic leaf of $\pi(G_r)$ and an embedding 
of $\pi(G_r)$ into a Lagrangian submanifold of $P$ defined by
\begin{eqnarray}
\hat g\ \mapsto\ ([1],\hat g).
\end{eqnarray}
Besides, $P$ is a groupoid with the multiplication
composing the $G$-orbits of the pairs 
$([\hat h],\hat g)$ and $([{\hat h}'],{\hat g}')$ such that 
$h_\pi\hat gh_\pi^{-1}=h'_0{\hat g}'{h'_0}^{-1}$ to the $G$-orbit of 
$([{\hat h}''],\hat g)$, where 
\begin{eqnarray}
{\hat h}''(x)=\cases{\hbox to 3.5cm{$\hat h(2x)$\hfill}{\rm for}
\quad 0\leq x\leq{1\over2}\pi\cr
\hbox to 3.5cm{${\hat h}'(2x-\pi){h'_0}^{-1}h_\pi$\hfill}{\rm for}
\quad{1\over2}\pi\leq x\leq\pi.}
\end{eqnarray}
The symplectic space $P$ together with the above projections,
the embedding and the partial multiplication satisfies
the axioms of the Poisson groupoid of $\pi(G_r)\cong G_r/G_0$
and provides an example of the general construction discussed
in \cite{cafe2}.

Coming back to the expression (\ref{Omega}) for the symplectic form 
$\widetilde\Omega$, we may simplify it using the freedom to choose 
a representative in the orbits of pairs $([\hat h],\hat g)$. 
Setting, for example, $h_0=1$, we obtain the formula
\begin{eqnarray}\label{fin}
\widetilde\Omega(\hat h,\hat g)={1\over4\pi}\Omega(h_\pi,\hat g)
\end{eqnarray}
where the last 2-form is given by the right hand side of (\ref{Omega2}).
As we see, the Poisson groupoid of $\pi(G_r)\cong G_r/G_0$ (as well
as that of $G_r$ itself) is closely related to the Heisenberg double 
of $G$. Its dimension is twice the dimension of $G$, as was indicated
by the general discussion. 

An alternative description of the symplectic structure on $P$ is 
obtained by taking representatives of (generic) orbits in the form
$(h_0,h_\pi,\ee^{2\pi\tau})$ with $\tau$ in the Cartan subalgebra
of $\ggl$ (this does not fix the ambiguity completely). In the above 
parametrization
\begin{eqnarray}\label{Omega3}
\widetilde\Omega=\Omega^{PL}(\tau,h_0)-\Omega^{PL}(\tau,h_\pi)
\end{eqnarray}
where 
\begin{eqnarray}\label{opl}
\Omega^{PL}(\tau,h)=\tr[d\tau\wedge(h^{-1}dh) + 
{1\over 4\pi}
(h^{-1}dh\wedge \ee^{2\pi\tau}h^{-1}dh\ \ee^{-2\pi\tau})]\,\cr
-{1\over 4\pi}\rho(h\ee^{2\pi\tau}h^{-1}).\hspace{1.5cm}
\end{eqnarray}
The form $\Omega^{PL}$ appeared for the first time in \cite{FG0}, see 
also \cite{AlMa} where its relation to the Heisenberg double has been 
stressed.

\nsection{Quantization}

Formula (\ref{Omega3}) should be compared to the one for the symplectic
form of the cotangent bundle $T^*G\cong G\times\ggl^*\cong G\times\ggl
\ni(g,p)$. Upon parametrization $h=h_\pi h_0^{-1}$, $p=h_0\tau h_0^{-1}$,
the canonical symplectic form on the cotangent
bundle $\Omega_{_{T^*G}}=d\,\tr[p(h^{-1}dh)]$ becomes
\begin{eqnarray}\label{oct}
\textstyle \Omega_{_{T^*G}}=\omega(\tau,h_\pi)-\omega(\tau,h_0)
\end{eqnarray}
where
\begin{eqnarray}\label{oco}
\omega(\tau,h)=\tr[d\tau\wedge (h^{-1}dh)-\tau(h^{-1}dh)^{\wedge 2}]=
\lim\limits_{k\to\infty}\ k\,\Omega^{PL}(\tau/k,h).
\end{eqnarray}
Let us recall the basic elements of the standard quantization 
of $T^*G$ adapted to the setup of complex manifolds. The space of 
quantum states $\CH$ and its dual $\CH^*$ are taken as
\begin{eqnarray}\label{L2}
\CH=\CA(G)\,,\qquad \CH^*=\prod\limits_\lambda End(V_\lambda)
\end{eqnarray} 
where $\CA(G)$ is the space of analytic functions on $G$ that
are finite linear combinations of matrix elements of irreducible analytic 
representation of $G$ of highest weight $\lambda$ acting in spaces 
$V_\lambda$ of dimension $d_\lambda<\infty$. 
The duality between $\CH$ and $\CH^*$ is given by
\begin{eqnarray}
\langle f,(u_\lambda)\rangle=\sum\limits_\lambda d_\lambda
\int f(g)\,\tr[g^{-1}_\lambda u_\lambda]dg
\end{eqnarray}
where the integral is over the compact form of $G$ and $g_\lambda
\in End(V_\lambda)$ represents $g$. Note that $\CH^*$, as
a direct product of matrix algebras, carries a natural algebra 
structure. The product in this algebra corresponds by the duality 
to the coproduct $f(g)\mapsto f(g_1g_2)$ in $\CA(G)$. Conversely, 
the commutative pointwise product of functions in $\CA(G)$ induces 
by the duality the coproduct in $\CH^*$. With the antipode defined 
as the transpose of the antipode $f(g)\mapsto f(g^{-1})$ of $\CA(G)$, 
$\CH^*$ becomes the Hopf algebra, the dual of $\CA(G)$. As such, 
it may be identified with a completion of the enveloping algebra 
$\CU(\ggl)$ of the Lie algebra $\ggl$ that embeds into it by
$\CU(\ggl)\ni u\mapsto(u_\lambda)\in\CH^* $ where $u_\lambda$ 
represents $u$ in $V_\lambda$. The algebra $\CU(\ggl)$ may be 
thought of as obtained by the quantization of the polynomial 
functions on $T^*G$ depending on the momentum $p=h_0\tau h_0^{-1}$. 
On the other hand, the quantization of the analytic functions 
on $T^*G$  that depend on $h=h_\pi h_0^{-1}$ (and Poisson-commute) 
gives rise to the commutative algebra $\CA(G)$ that we identified
with $\CH$.

In the deformed case, one may take as the dual space of boundary 
states
\begin{eqnarray}\label{L3}
\CH^*_q=\prod\limits_\lambda End(V^q_\lambda)
\end{eqnarray} 
where $V^q_\lambda$ are the highest weight representations of
the quantum deformation $\CU_q(\ggl)$ of the enveloping algebra
$\CU(\ggl)$ corresponding to the classical r-matrix $r$, whenever
those are available. Again $\CH^*$ has a natural structure of algebra
and may be thought of for generic $q$ (related to the coupling 
constant $k$) as a completion of $\,\CU_q(\ggl)$ and, also, as a 
quantization of the space of functions on $P$ that depend on 
the variable $h_0\,\ee^{2\pi\tau}h_0^{-1}$. Similarly, for generic 
$q$, the classical function on $P$ that depend on $h=h_\pi h_0^{-1}$ 
give rise upon quantization to the (non-commutative) algebra $\CF_q(G)$
of function on the quantum group that may be identified with 
the space of states $\CH_q$ \cite{S-T-S1}. The duality interchanges 
products and coproducts. The algebraic aspects of the quantization 
of the Heisenberg double were also studied in \cite{AlFa}

When $q$ is a root of unity then the algebra $\,\CU_q(\ggl)$
has a finite-dimensional quotient $\widetilde{\CU}_q(\ggl)$ 
\cite{Lusztig} which has a finite series of irreducible 
representations labeled by the so called integrable weights 
$\lambda$ which may be used in (\ref{L3}). The finite-dimensional 
boundary algebra $\CH_q^*$ obtained this way may be equipped with 
the non-degenerate bilinear form given by $\pm S^\lambda_0\,\tr$ 
on each component $End(V^q_\lambda)$, where $S^\lambda_\zeta$ is 
the modular matrix. Together with the fusion algebra of bulk states, 
one obtains then a genuine example of a topological two-dimensional 
field theory in the sense of \cite{Laza}, see also \cite{GawGH}. 
It may be thought of as a ``vertex version'' of the boundary $G/G$ 
theory studied in \cite{GawGH}. In this case, however, the algebra 
$\CH_q^*$ does not carry a natural coproduct (the map 
$\widetilde{\CU}_q(\ggl)\ni u\mapsto(u_\lambda)\in\CH_q^*$ is not 
an isomorphism and it may not be used to transport the coproduct 
from $\,\widetilde{\CU}_q(\ggl)$).

\end{document}